\documentclass[a4paper,12pt]{article}  
\usepackage{amsfonts} 
\usepackage{amsmath} 
\usepackage{ifthen}
\numberwithin{equation}{section}
\begin{document} 
%+++++++++++++++++++++++++++++++++++++++++++++++++++++++++++++++++++++ 
\title{Finite--amplitude inhomogeneous plane waves \\ 
       in a deformed Mooney--Rivlin material.} 
%+++++++++++++++++++++++++++++++++++++++++++++++++++++++++++++++++++++ 
 
\author{Michel Destrade} 
\date{2000} 
\maketitle 
 
 \bigskip
 
%++++++++++++++++++++++++++++++++++++++++++++++++++++++++++ 

\begin{abstract} 

The propagation of finite--amplitude linearly--polarized inhomogeneous 
transverse plane waves is considered for a Mooney-Rivlin material 
maintained in a state of finite static homogeneous deformation. 
It is shown that such waves are possible provided that the directions 
of the normal to the planes of constant phase and of the normal to 
the planes of constant amplitude are orthogonal and conjugate with 
respect to the \textbf{B}--ellipsoid, where \textbf{B} is the left 
Cauchy-Green strain tensor corresponding to the initial deformation. 
For these waves, it is found that even though the system is
non-linear, results on energy flux are nevertheless identical with 
corresponding results in the classical linearized elasticity theory. 
Byproducts of the results are new exact static solutions for 
the Mooney-Rivlin material. 

\end{abstract} 

%++++++++++++++++++++++++++++++++++++++++++++++++++++++++++  

\newpage

%++++++++++++++++++++++++++++++++++++++++++++++++++++++++++ 
\section{Introduction} 
%++++++++++++++++++++++++++++++++++++++++++++++++++++++++++ 
 
An isotropic homogeneous elastic material maintained in a state of
arbitrary finite static homogeneous deformation exhibits three
privileged orthogonal directions, namely those of the principal axes
of strain. 
For a general material held in such a state, Green
\cite{Gree63} showed that transverse finite--amplitude homogeneous
plane waves can travel in a principal direction and \mbox{Carroll}
\cite{Carr67} showed that homogeneous transverse plane waves of finite
amplitude, linearly--polarized in a principal direction, may 
propagate in a principal plane.
 
John \cite{John66} considered the possibility of having 
finite--amplitude plane waves propagating in
\textit{any} direction in a prestressed material; 
he showed that the so-called Hadamard material
was the most general compressible one in which it may happen. 
Then Currie and Hayes \cite{CuHa69} extended his result to 
incompressible bodies and the corresponding material was found to be 
of the Mooney--Rivlin type, a model used to describe the mechanical 
behaviour of rubber \cite{Moon40}. 
Later, Boulanger and Hayes \cite{BoHa92,BoHa95} gave a detailed study 
of finite--amplitude waves propagating in a deformed Mooney--Rivlin
material.
For these waves, the displacement is of the form 
$g(\mathbf{n\cdot x}-vt)\mathbf{a}$, 
where $g$ is a function of arbitrary 
magnitude, \textbf{n} is a unit vector in the direction of propagation
(which may be any direction), $v$ is the speed at which the wave 
travels and \textbf{a} is a unit vector in the direction of 
polarization. 
 
In contrast to this type of wave, \textit{inhomogeneous} plane waves 
have distinct planes of constant amplitude and of constant phase. 
The purpose of this paper is to consider the superposition of an 
inhomogeneous motion upon an arbitrary static homogeneous deformation 
in a Mooney--Rivlin material. 
We show that waves with displacement of the form
$f(\mathbf{b\cdot x})g(\mathbf{n\cdot x}-vt)\mathbf{a}$, 
where $f$ and $g$ are
certain functions of finite magnitude, and the vectors
\textbf{n} and \textbf{b} satisfy a geometric relationship, may
propagate in  the deformed Mooney--Rivlin material,
for \textit{any} orientation of the plane of \textbf{b} and \textbf{n}.

The plan of the paper is as follows.  
 
In Section \ref{Basic_equations}, 
we recall the basic equations describing the Mooney--Rivlin 
incompressible material.

In Section \ref{Motion_superposed_on_static_homogeneous_deformation},
we assume that this material is subjected to a finite homogeneous
static deformation, upon which a finite--amplitude 
linearly--polarized inhomogeneous plane wave is superposed. 

The equations of motion are derived in 
Section \ref{Equations_of_motion} and solved for the
functions $f$ and $g$, the speed $v$ and the pressure. 
Two types of solutions arise.
When the directions of \textbf{a}, \textbf{b}, and \textbf{n}
are along principal
axes of the initial strain ellipsoid, the solutions are called
`special principal motions' (Section \ref{Special_principal_motions});
when the orientation of the plane containing  
\textbf{b} and \textbf{n}  is arbitrary,
we show that these directions must be 
conjugate with respect to the elliptical section of the strain 
ellipsoid corresponding to the initial static deformation by the plane
orthogonal to the direction of polarization 
(Section \ref{General_solutions}).

In Section \ref{Sinusoidal_evanescent_waves}, 
we study a class of solutions which are
of interest in some physical contexts (gravity waves, surface waves, 
 waves in layered media, interfacial waves, etc\ldots). 
For these solutions, the amplitude decays
exponentionally in one direction while it varies  sinusoidally with 
time in another. 
We show how the possible directions of polarization, propagation and 
attenuation may be constructed geometrically, and give bounds for the 
values of the phase speed. 
For these waves, the propagation of energy is also considered. 
Although the theory is non-linear, well-known aspects of the 
linearized theory of wave propagation in conservative media are
recovered. 
In particular, it is shown that the direction of 
the mean energy flux vector is parallel 
to the planes of constant amplitude,
and that the component of this vector in the direction of propagation
is equal to the phase speed times the total energy 
density. 

Finally, in Section \ref{Superposed_static_deformation},
we investigate the case where the time 
dependency of the superposed deformation is removed. 
The corresponding solutions provide examples of inhomogeneous 
finite static deformations possible in a Mooney--Rivlin material, 
in the absence of body forces.

%++++++++++++++++++++++++++++++++++++++++++++++++++++++++++++++++++++ 
\section{Basic equations}
\label{Basic_equations} 
%++++++++++++++++++++++++++++++++++++++++++++++++++++++++++++++++++++ 
 
Here we recall the basic equations governing the deformation of 
Mooney--Rivlin rubberlike materials.

The constitutive equation for a Mooney--Rivlin material is 
\begin{equation} \label{T} 
\mathbf{T} = -p\mathbf{1}+(C+DI)\mathbf{B}-D \mathbf{B}^2
=-(p- DII)\mathbf{1}+C \mathbf{B}-D \mathbf{B}^{-1}. 
\end{equation} 
Here  $\mathbf{T}$ is the Cauchy stress,  $C$ and $D$ are constants,
 $\mathbf{B}$ is the left Cauchy-Green strain tensor defined by   
$\mathbf{B}=\mathbf{F}\mathbf{F}^{\mathrm{T}}$,
where $\mathbf{F}=\partial\mathbf{x} / \partial\mathbf{X}$ is the  
deformation gradient, and the deformation is 
$\mathbf{x}= \mathbf{x}(\mathbf{X},t)$.
Also, $I$ and $II$ are invariants of \textbf{B}, given by
$I=\rm{tr} \, \mathbf{B},  \quad 
\textit{II}= \rm{tr} \, \mathbf{B}^{-1}$. 

In the constitutive equation \eqref{T}, $p(\mathbf{x},t)$ 
corresponds to an arbitrary pressure. 
Because the material is incompressible, we must have 
\begin{equation} \label{inc} 
\mathrm{det}\, \mathbf{F}=1.
\end{equation} 

For this material, the strain energy density $\Sigma$,
measured per unit volume in the current state of deformation, 
has the form \cite{Moon40}   
\begin{equation} \label{sigma} 
2\Sigma=C(I-3)+D(II-3).
\end{equation} 
It is assumed that 
\begin{equation} \label{ellip} 
C \geq 0, \,  D>0, \qquad 
\mathrm{or} \qquad
C>0, D \geq 0, 
\end{equation} 
in order that the strong ellipticity condition be  satisfied 
(see \cite{BoHa95} for a short proof). 
If $D=0$, the material is said to be neo--Hookean which case is 
being examined elsewhere.
Here we exclude this possibility, that is we assume that
 \eqref{ellip}$_1$ holds.
  
The equations of motion in the absence of body forces are 
\begin{equation} \label{Motiondef} 
\mathrm{div} \, \mathbf{T} =
\rho\frac{\partial ^2 \mathbf{x}}{\partial t^2},
\quad
\frac{\partial T_{ij}}
{\partial x_j}
 = \rho  \frac{\partial^2 x_i}{\partial t^2},
\end{equation} 
where $\rho$ is the mass density of the material, measured 
per unit volume in any configuration 
(because of the incompressibility constraint).

The energy flux vector $\mathbf{R}$
is defined by 
\begin{equation} \label{Rdef}
\mathbf{R} = - \mathbf{T}\cdot \dot{\mathbf{x}},
\quad 
R_i = - T_{ij} \, \dot{x}_j,
\end{equation}
where $ \dot{\mathbf{x}}$ is the particle velocity.
$R_{\alpha}$ gives the rate at which the mechanical  energy 
crosses, at  time $t$,  a material element which is normal to the 
$x_{\alpha}$--axis at  time $t$, in the final state of deformation,
measured per unit area of surface in this configuration.

Also, the kinetic  energy density $K$ measured per unit volume  
is given by 
\begin{equation} \label{Kdef}
K =  \rho  (\dot{\mathbf{x}}\cdot \dot{\mathbf{x}})/2.
\end{equation}

%++++++++++++++++++++++++++++++++++++++++++++++++++++++++++++++
\section{Motion superposed on static homogeneous deformation} 
\label{Motion_superposed_on_static_homogeneous_deformation}
%++++++++++++++++++++++++++++++++++++++++++++++++++++++++++++++
 
Here we consider the propagation of a linearly--polarized 
inhomogeneous plane wave of finite amplitude in the material,
when it is held in a state of finite static homogeneous deformation.
We determine the corresponding stresses and energy flux for the 
motion.

We assume that the material is held in the state of finite static 
homogeneous deformation given by 
\begin{equation}
x_{\alpha}= \lambda_{\alpha} X_{\alpha}, \quad 
\alpha = 1,2,3,
\end{equation}
in which the particle initially at position $\mathbf{X}$ is 
displaced to $\mathbf{x}$. 
The constants $\lambda_{1},\lambda_{2},\lambda_{3}$ are such that
$\lambda_{1} > \lambda_{2} > \lambda_{3} >0$ and satisfy 
$\lambda_{1} \lambda_{2} \lambda_{3}=1 $, due to \eqref{inc}.

In this case, the principal stresses $t_{\alpha}$, necessary to
support the deformation are given by
\begin{equation}
t_{\alpha}= -p_0 + (C+D I) \lambda_{\alpha}^2 - D\lambda_{\alpha}^4,
\quad \alpha = 1,2,3,
\end{equation}
where now $I=  \lambda_1^2 + \lambda_2^2 + \lambda_3^2$ and $p_0$
is a constant.

Let a linearly--polarized inhomogeneous 
 plane wave of finite amplitude propagate in the deformed body, so 
that the final position of the particle, which is initially at 
\textbf{X}, and at \textbf{x} in the state of finite static 
homogeneous deformation, is at $\overline{\mathbf{x}}$ where 
\begin{equation} \label{xBar} 
\overline{\mathbf{x}} = \mathbf{x}+ 
f(\mathbf{b\cdot x})g(\mathbf{n\cdot x}-vt) \mathbf{a}. 
\end{equation} 
Here, (\textbf{n}, \textbf{a}, \textbf{b}) form an orthonormal triad. 
The planes defined by $\mathbf{n\cdot x}= const.$ are the planes of 
constant phase and those defined by $\mathbf{b\cdot x}= const.$ 
are the planes of constant amplitude. 
Also, $v$ is the real speed of propagation and ($f, g$) are two 
real functions to be determined. 

We use bars, for example $\overline{W}$, to denote quantities in the 
final state of deformation. 
The deformation gradient $\overline{\mathbf{F}}$ associated with 
the deformation \eqref{xBar} is given by  
\begin{equation} \label{Fbar} 
\overline{\mathbf{F}} = 
\frac{\partial\overline{\mathbf{x}}}{\partial\mathbf{X}}= 
[\mathbf{1} 
+f'g \, \mathbf{a \otimes b}  
+fg' \, \mathbf{a \otimes n}]\mathbf{F}, 
\end{equation} 
where the prime denotes the derivative of a function with respect 
to its argument. 
 
The corresponding left Cauchy-Green tensor  
$\overline{\mathbf{B}}=\overline{\mathbf{F}} \, 
\overline{\mathbf{F}}^{\mathrm{T}}$ 
and its inverse 
$\overline{\mathbf{B}}^{-1}= 
\overline{\mathbf{F}}^{-\mathrm{T}} \, 
\overline{\mathbf{F}}^{-1}$ are 
\begin{equation} \label{Bbar} 
\begin{array}{l} 
\overline{\mathbf{B}} =  
[\mathbf{1} 
+f'g \, \mathbf{a \otimes b}  
+fg' \, \mathbf{a \otimes n}]\mathbf{B}  
[\mathbf{1} 
+f'g \, \mathbf{b \otimes a}  
+fg'  \,\mathbf{n \otimes a}] ,\\ 
\overline{\mathbf{B}}^{-1} =  
[\mathbf{1} 
-f'g  \, \mathbf{b \otimes a}  
-fg'  \, \mathbf{n \otimes a}]\mathbf{B}^{-1}  
[\mathbf{1} 
-f'g  \, \mathbf{a \otimes b}  
-fg'  \, \mathbf{a \otimes n}]. 
\end{array} 
\end{equation} 
Also, the invariants of $\overline{\mathbf{B}}$
 are given by
\begin{equation} \label{inv} 
\begin{split} 
\overline{I}  &= \mathrm{tr} \, \overline{\mathbf{B}} \\
& = I +2[f'g (\mathbf{a \cdot Bb}) + fg' (\mathbf{a \cdot Bn})] \\ 
& \phantom{=  I +2[} 
+ (f'g )^2 (\mathbf{b \cdot Bb})+2ff'gg' (\mathbf{n \cdot Bb}) 
+( fg' )^2 (\mathbf{n \cdot Bn}), \\ 
\overline{II} &= \mathrm{tr} \, \overline{\mathbf{B}}^{-1} \\ 
&=II - 2[f'g (\mathbf{a \cdot B^{-1} b}) 
+fg' (\mathbf{a \cdot B^{-1} n})] 
+[( f'g )^2 +( fg' )^2](\mathbf{a \cdot B^{-1}a}). 
\end{split}
\end{equation} 
 
We introduce the coordinates ($\eta, \xi, \zeta$), given by 
\begin{equation} \label{coord} 
\eta=\mathbf{n\cdot x}, \quad 
\xi=\mathbf{a\cdot x}, \quad 
\zeta=\mathbf{b\cdot x}, 
\end{equation} 
and the coordinates 
($\overline{\eta}, \overline{ \xi}, \overline{ \zeta}$), given by 
\begin{equation} \label{coordBar} 
\overline{\eta} =\mathbf{n\cdot  \overline{x}}= \eta , \quad 
\overline{\xi}=\mathbf{a\cdot  \overline{x}}=\xi 
                + f(\zeta)g(\eta - vt), \quad  
\overline{\zeta}=\mathbf{b\cdot \overline{x}}=\zeta. 
\end{equation} 

Now we  write the Cauchy stress tensor $\overline{\mathbf{T}}$ 
in (\textbf{n}, \textbf{a}, \textbf{b}). 
For a Mooney--Rivlin material, the stress-strain relation is 
given by equation \eqref{T} or, in the present context, by 
\begin{equation} \label{Tbar} 
\overline{\mathbf{T}} = -(\overline{p} - D \overline{II}) \mathbf{1}
 +C \overline{\mathbf{B}} - D \overline{\mathbf{B}}^{-1}, 
\end{equation} 
where the arbitrary pressure $\overline{p}$ may be decomposed 
into the pressure $p$ corresponding to the primary homogeneous 
deformation and an incremental pressure $p^*$ (say) corresponding 
to the superposed dynamic deformation: $\overline{p}=p+p^*$. 
We assume $p^*$ to be of a form similar to the superposed motion 
\eqref{xBar}, that is 
$p^* = p^*(\mathbf{n\cdot x}, \mathbf{b\cdot x}, t) 
= p^*(\eta, \zeta, t)$. 
 
Using  \eqref{Bbar}, \eqref{inv}, \eqref{Tbar},  
and the  notation:  
\mbox{$\overline{\mathbf{T}}_{\eta\eta}= 
\mathbf{n \cdot \overline{T} n}$}, 
\mbox{$\overline{\mathbf{T}}_{\eta\xi}=  
\mathbf{n \cdot \overline{T} a}$}, etc.,  
the components of $\overline{\mathbf{T}}$  are found as  
\begin{equation} \label{TComp} 
\begin{array}{l}
\overline{\mathbf{T}}_{\eta\eta}= \mathbf{T}_{\eta\eta}-p^* 
-2D f'g (\mathbf{a \cdot B^{-1} b}) 
+D( f'g )^2 (\mathbf{a \cdot B^{-1} a}),      \\ 
\overline{\mathbf{T}}_{\eta\xi}= \mathbf{T}_{\eta\xi} 
+C[ fg' (\mathbf{n \cdot B n}) 
+f'g (\mathbf{n \cdot B b})] 
+D (fg')^2 (\mathbf{a \cdot  B^{-1} a}),            \\ 
\overline{\mathbf{T}}_{\eta\zeta}=\mathbf{T}_{\eta\zeta}
+D [f'g (\mathbf{a \cdot  B^{-1} n}) 
+fg' (\mathbf{a \cdot  B^{-1} b})] 
-D ff'gg' (\mathbf{a \cdot  B^{-1} a}),      \\ 
\begin{split} 
\overline{\mathbf{T}}_{\xi\xi}= \mathbf{T}_{\xi\xi}-p^* 
+ 2 & C[ f'g (\mathbf{a \cdot B b} ) 
+ fg' (\mathbf{a \cdot Bn}) ] \\ 
+ & C[( f'g )^2 (\mathbf{b \cdot B b}) 
+2ff'gg' (\mathbf{n \cdot B b} ) 
+( fg' )^2 (\mathbf{n \cdot B n})], 
\end{split} \\ 
\overline{\mathbf{T}}_{\xi\zeta}= \mathbf{T}_{\xi\zeta}
+C[ f'g (\mathbf{b \cdot Bb}) 
+fg' (\mathbf{n \cdot Bb})] 
+D f'g (\mathbf{a \cdot  B^{-1} a}),               \\ 
\overline{\mathbf{T}}_{\zeta\zeta}= \mathbf{T}_{\zeta\zeta}-p^* 
-2D fg'(\mathbf{a \cdot  B^{-1} n}) 
+D(fg')^2(\mathbf{a\cdot  B^{-1} a}).   
\end{array} 
\end{equation} 

In passing, we introduce some quantities associated with the 
energy carried by the disturbance \eqref{xBar}.
First, from \eqref{Kdef} and  \eqref{xBar}, the kinetic energy density 
$\overline{K}$ per unit volume is  given by
\begin{equation} \label{kinetic_energy}
\overline{K} =  \rho  v^2 (fg')^2 /2.
\end{equation} 

Next, the stored--energy density $\overline{W}$ per unit volume
associated with the wave is defined, in the absence of body forces, by 
\begin{equation}  \label{stored_energy}
\overline{W} = \overline{\Sigma}- \Sigma  
=C(\overline{I}-I)/2 + D (\overline{II}-II)/2, 
\end{equation} 
where $\overline{I}$ and $\overline{II}$ are given by equations
\eqref{inv} in our  context.

Finally, we introduce the energy flux vector $\overline{\mathbf{R}}$
associated with the motion \eqref{xBar}. 
From \eqref{Rdef} it is given here by
$\overline{\mathbf{R}} 
= -\overline{\mathbf{T}}\cdot \dot{\overline{\mathbf{x}}}$. 
It is related to the energy flux vector \textbf{R} measured 
in the intermediate static state of deformation through \cite{HaRi72} 
\begin{equation} 
\mathbf{R}= \big{(} \frac{\partial \overline{\mathbf{x}}} 
{\partial \mathbf{x}}\big{)}^{-1} \overline{\mathbf{R}}  
=- \big{(} \frac{\partial \overline{\mathbf{x}}} 
{\partial \mathbf{x}}\big{)}^{-1}  \overline{\mathbf{T}} 
\cdot \dot{\overline{\mathbf{x}}} 
\end{equation} 
Using \eqref{xBar} we have
$\partial \overline{\mathbf{x}} / \partial \mathbf{x}
=\mathbf{1} + f'g \mathbf{a} \otimes \mathbf{b} 
+ fg' \mathbf{a} \otimes \mathbf{n}$, so that
\begin{equation} \label{energy_flux}
\mathbf{R}= v (fg')
[\mathbf{1} - f'g \mathbf{a} \otimes \mathbf{b} 
- fg' \mathbf{a} \otimes \mathbf{n}]
 \overline{\mathbf{T}} \cdot \mathbf{a}.
\end{equation}

%++++++++++++++++++++++++++++++++++++++++++++++++++
\section{Equations of motion} 
\label{Equations_of_motion}
%++++++++++++++++++++++++++++++++++++++++++++++++++
 
The equations of motion \eqref{Motiondef} are written as  
\begin{equation} \label{mtn1} 
\mathrm{div} \, \overline{\mathbf{T}} =
\rho\frac{\partial ^2 \overline{\mathbf{x}}}{\partial t^2},
\quad
\frac{\partial \overline{\mathbf{T}}_{ij}}
{\partial \overline{x}_j}
 = \rho  \frac{\partial^2 \overline{x}_i}{\partial t^2}
\end{equation} 
Here $\mathrm{div}$ 
represents the divergence operator with respect to position 
$\overline{\mathbf{x}}$, that is with respect to  
the coordinates ($\overline{\eta}, \overline{ \xi}, \overline{ \zeta})
 =(\eta, \overline{ \xi}, \zeta$).
However, by inspection of equations \eqref{TComp}, we see that  
$\overline{\mathbf{T}}$ depends on $\eta, \zeta$ and $t$ only, so that 
$\mathrm{div} \, \overline{\mathbf{T}}$ computed with respect to  
$\overline{\mathbf{x}}$ is equal to 
$\mathrm{div} \, \overline{\mathbf{T}}$ computed with respect to 
$\mathbf{x}$. 
Hence the equations of motion reduce to 
\begin{equation} 
\left. 
\begin{array}{l} 
\overline{\mathbf{T}}_{\eta\eta,\eta}  
+\overline{\mathbf{T}}_{\eta\zeta,\zeta}=0,              \\ 
\overline{\mathbf{T}}_{\xi\eta,\eta}  
+\overline{\mathbf{T}}_{\xi\zeta,\zeta}=\rho v^2 f g'',\\ 
\overline{\mathbf{T}}_{\zeta\eta,\eta}  
+\overline{\mathbf{T}}_{\zeta\zeta,\zeta}=0,    
\end{array} 
\right\}
\end{equation} 
or  
\begin{equation} \label{mtn2} 
\left.
\begin{split} 
-p^*_{,\eta} 
+D [f''g (\mathbf{a \cdot B^{-1}n}) - &
f'g' (\mathbf{a \cdot B^{-1}b})]  \\ 
 & +D({f'}^2-ff'') gg'(\mathbf{a \cdot B^{-1}a})=0, \\ 
C [f''g (\mathbf{b \cdot Bb})  
+ 2 f'g' & (\mathbf{n \cdot Bb}) 
+ fg'' (\mathbf{n \cdot Bn})] \\ 
 & +D(f''g+fg'') (\mathbf{a \cdot B^{-1}a}) 
=\rho v^2 f g'', \\
-p^*_{,\zeta} 
+D [fg'' (\mathbf{a \cdot B^{-1}b}) - &
f'g' (\mathbf{a \cdot B^{-1}n})] \\
& +D({g'}^2-gg'') ff'(\mathbf{a \cdot B^{-1}a})=0, 
\end{split} 
\right\} 
\end{equation} 
where commas denote differentiation with respect to the coordinates 
($\eta, \xi, \zeta$); thus, for example,  
$\overline{\mathbf{T}}_{\zeta\xi,\xi}=  
\partial \overline{\mathbf{T}}_{\zeta\xi}/\partial \xi$. 
 
Now, equation $\eqref{mtn2}_{2}$ is equivalent to 
\begin{multline} \label{2ndOrder} 
[C (\mathbf{b\cdot Bb})  
+D(\mathbf{a \cdot B^{-1}a})]f''g  
+ 2 C(\mathbf{n\cdot Bb}) f'g' \\ 
+[C(\mathbf{n\cdot Bn}) 
+D(\mathbf{a \cdot B^{-1}a}) 
-\rho v^2] fg'' = 0. 
\end{multline} 
 
Also, the second derivatives of $p^*$ must be compatible, that is 
\mbox{$p^*_{,\eta\zeta}=p^*_{,\zeta\eta}$}, or, from
$\eqref{mtn2}_{1,3}$,
\begin{multline} \label{compat} 
f'''g (\mathbf{a \cdot B^{-1}n}) 
-f''g'(\mathbf{a \cdot B^{-1}b}) 
+(f'f''-ff''')gg'(\mathbf{a \cdot B^{-1}a}) 
\\ 
=fg''' (\mathbf{a \cdot B^{-1}b}) 
-f'g''(\mathbf{a \cdot B^{-1}n}) 
+(g'g''-gg''')ff'(\mathbf{a \cdot B^{-1}a}). 
\end{multline} 

Equations \eqref{2ndOrder} and \eqref{compat} are the two equations
to be solved for $f$ and $g$ in order that the inhomogeneous 
motion \eqref{xBar} may propagate in the deformed Mooney--Rivlin
material.
The solutions are established in Appendix A, 
and a distinction needs be made, according as to whether or not
 the orthogonal vectors \textbf{n}, \textbf{a}, and \textbf{b} are 
along the principal axes of the \textbf{B}--ellipsoid.

%++++++++++++++++++++++++++++++++++++++++
\section{Special principal motions} 
\label{Special_principal_motions}
%++++++++++++++++++++++++++++++++++++++++
 
Here we present solutions valid  when  \textbf{n}, \textbf{a}, 
and \textbf{b} are in the principal directions of the initial strain
ellipsoid.
 
For small-amplitude homogeneous plane waves propagating in a
homogeneously deformed body, the term `principal wave' \cite{Mana59}
is used to describe a wave travelling along a principal axis of the 
strain ellipsoid corresponding to the initial static deformation.
This terminology can readily be extended to the case of inhomogeneous 
motions for which the planes of constant phase are orthogonal to such 
an axis.

We introduce the term `special principal motion' to describe an 
inhomogeneous motion for which the normal to the planes of constant 
phase,  the normal to the planes of constant amplitude, 
and the linear polarization are in the directions of 
the  principal axes of strain. 
With our notation, a special principal motion is of the form 
$f(\mathbf{b\cdot x})g(\mathbf{n\cdot x}-vt) \mathbf{a}$ where 
\textbf{n}, \textbf{a}, and \textbf{b} are the vectors 
\textbf{i}, \textbf{j}, and \textbf{k}.

The most general functions $f$ and $g$, solutions to the equations of
motion \eqref{2ndOrder} and \eqref{compat}, where, without loss of 
generality, $\mathbf{n}=\mathbf{i}$, $\mathbf{a}=\mathbf{j}$, and 
$\mathbf{b}=\mathbf{k}$, 
are such that one is of exponential type while the 
other is of sinusoidal type (see Proof in Appendix A). 
For these motions, the quantity $v$ 
may  be arbitrarily prescribed (within an interval).

Explicitly, the possible inhomogeneous special principal motions 
of the Mooney--Rivlin material are written as
\begin{equation} \label{SpecialMtn}
\overline{x}= \lambda_1 X, \quad
\overline{y}= \lambda_2 Y +f(\lambda_3 Z) g(\lambda_1 X -vt), \quad  
\overline{z}= \lambda_3 Z , 
\end{equation}
where either
\begin{equation} 
\left.
\begin{array}{l}
f(\lambda_3 Z)= f(z) =
a_1 e^{ k \sigma z} +a_2 e^{-k \sigma z}, \\
g(\lambda_1 X -vt)=  g(x-vt) =
d_1 \cos k (x-vt) + d_2 \sin k(x -vt),
\end{array}
\right\}
\end{equation} 
or 
\begin{equation} 
\left.
\begin{array}{l}
f(\lambda_3 Z)= f(z) =
a_1 \cos k \sigma z
 +a_2 \sin k  \sigma z, \\
g(\lambda_1 X-vt)=  g( x-vt) =
 d_1 e^{ k (x-vt)} + d_2 e^{ - k(x -vt)}.
\end{array}
\right\}
\end{equation} 
Here, $a_1, a_2, d_1, d_2$ are constants, $\sigma$ is defined 
by
\begin{equation}
\sigma =
\sqrt{\frac{C \lambda_3^2 +D \lambda_2^{-2} - v^2}
 {C \lambda_1^2 +D \lambda_2^{-2}}},  
\end{equation} 
and $k$ and $v$ are arbitrary 
$(0 \leq v^2 <C \lambda_3^2 +D \lambda_2^{-2})$.

%+++++++++++++++++++++++++++++++++++++++++++++++++++++++++
\section{General solutions} 
\label{General_solutions}
%++++++++++++++++++++++++++++++++++++++++++++++++++++++++
  
Here, we seek the most general solutions $f$ and $g$ 
of equation \eqref{2ndOrder} 
satisfying \eqref{compat} such that 
$f(\mathbf{b\cdot x}) g(\mathbf{n\cdot x}-vt) \mathbf{a}$ is an
\textit{inhomogeneous} motion in the deformed Mooney--Rivlin
material, (\textbf{a}, \textbf{n}, \textbf{b}) form an orthonormal 
triad, and the plane of \textbf{n} and \textbf{b} is  
\textit{arbitrary}. 
It is shown in  Appendix A 
that $f$ and $g$ are the functions defined by either
\begin{equation} \label{fg} 
\left.
\begin{array}{l}
f(\zeta)=a_1 e^{k \zeta}
 +a_2 e^{-k \zeta}, \\
g(\eta-vt)= d_1 \cos k (\eta-vt) + d_2 \sin k(\eta -vt),
\end{array}
\right\}
\end{equation} 
or
\begin{equation} \label{fgBis}
\left.
\begin{array}{l}
f(\zeta)=a_1 \cos k \zeta
 +a_2 \sin k \zeta, \\
g(\eta-vt)= d_1 e^{k (\eta-vt)} + d_2 e^ {- k(\eta -vt)}.
\end{array}
\right\}
\end{equation}
Here $a_1, a_2, d_1, d_2$ and $k$ are arbitrary constants, $v$  
is such that 
\begin{equation} \label{speed} 
\rho v^2 = C[(\mathbf{n \cdot Bn} ) -(\mathbf{b \cdot Bb} )], 
\end{equation} 
and the condition 
\begin{equation} \label{conj} 
\mathbf{b \cdot Bn}=0, 
\end{equation} 
must be satisfied.  
 
The condition \eqref{conj} means that the unit 
vectors \textbf{n} and \textbf{b} are conjugate with respect
to the \textbf{B}--ellipsoid, defined by $\mathbf{x\cdot Bx}=1$.
Because they are orthogonal, they must lie along the principal axes 
of the elliptical section of the \textbf{B}--ellipsoid by the 
plane orthogonal to \textbf{a}. 
We note that  the speed $v$ given by \eqref{speed} is real
when  $\mathbf{\hat{m}}$ and $\mathbf{\hat{n}}$  are in the
directions of the minor and major axes of the elliptical
section, respectively.
 
Also note that 
\begin{equation} 
\rho v^2 = \mathbf{n \cdot Tn} - \mathbf{b \cdot Tb} 
-\frac{\mathbf{b \cdot Tn}}{\mathbf{b \cdot B^{-1} n}}
(\mathbf{n \cdot B^{-1}n} -\mathbf{b \cdot B^{-1} b}), 
\end{equation} 
so that the speed may be written in terms of the basic strain 
\textbf{B} and the corresponding basic stress \textbf{T}.

Finally, using the compatibility equations, the 
incremental pressure $p^* $ is determined. 
To within an arbitrary constant, it is found to be either
\begin{multline} 
p^* = -D[fg'(\mathbf{a\cdot B^{-1}n}) 
+ f'g(\mathbf{a\cdot B^{-1}b})] \\
+\frac{D k^2}{2}[(d_1^2 + d_2^2) f^2 - 4 a_1 a_2 g^2] 
(\mathbf{a \cdot B^{-1} a}),
\end{multline} 
when $f$ and $g$ are given by \eqref{fg}, or
\begin{multline}
p^* =-D[fg'(\mathbf{a\cdot B^{-1}n}) + f'g(\mathbf{a\cdot B^{-1}b})] \\
+\frac{D k^2}{2}[4d_1^2 d_2^2 f^2 - ( a_1^2 + a_2^2) g^2] 
(\mathbf{a \cdot B^{-1} a}),
\end{multline} 
when $f$ and $g$ are given by \eqref{fgBis}.
Thus, a \textit{finite} displacement of the form 
$f(\mathbf{b\cdot x})g(\mathbf{n\cdot x}-vt) \mathbf{a}$, 
where (\textbf{a}, \textbf{b}, \textbf{n}) is an orthonormal basis, 
$f$ and $g$ are given by \eqref{fg} or \eqref{fgBis}, 
$v$ is given by \eqref{speed}, 
and \textbf{b}, \textbf{n} satisfy \eqref{conj}, is an exact solution 
to the equations of motion in a homogeneously deformed Mooney--Rivlin 
material, for any orientation of the plane of 
\textbf{b} and \textbf{n}.

%+++++++++++++++++++++++++++++++++++++++++++++++++++++++++++++++++++ 
\section{Sinusoidal evanescent waves} 
\label{Sinusoidal_evanescent_waves} 
%+++++++++++++++++++++++++++++++++++++++++++++++++++++++++++++++++++ 
  
In this Section, we restrict our attention to the propagation of a 
linearly--polarized finite--amplitude inhomogeneous plane wave in a
homogeneously deformed Mooney--Rivlin material. 
The phase of the wave fluctuates sinusoidally in the direction of a 
unit vector \textbf{n}, its amplitude decreases exponentionally in the
direction of \textbf{b}, orthogonal to \textbf{n}, and its 
polarization is in the direction of \textbf{a}, orthogonal to both 
\textbf{n} and \textbf{b}.  
These waves are a subclass of the general solutions to the equations of
motion found in the previous Section, and arise in a variety of 
contexts such as Raleigh waves, Love waves, Stoneley waves, etc \ldots

We give a geometrical construction for the triads 
(\textbf{n}, \textbf{a}, \textbf{b}) such that the wave may 
propagate, and present expressions for the phase speed and the 
pressure. 
Then we examine the propagation of the energy carried by the wave. 
 
%---------------------------------------------- 
\subsection{Construction} 
%---------------------------------------------- 
 
Henceforth, we consider the motion  
\begin{equation} \label{expcos} 
\overline{\mathbf{x}} = \mathbf{x}+  
\alpha e^{-k \mathbf{b\cdot x}} 
\cos k (\mathbf{n\cdot x}-vt) \mathbf{a}, 
\end{equation} 
where $\alpha$ and $k$ are real arbitrary constants, and $v$ is 
assumed to be real. 
Recall that in equation \eqref{expcos}, \textbf{x} corresponds
to the static homogeneous predeformation
$\mathbf{x}  = \mathrm{diag}(\lambda_1,\lambda_2,\lambda_3)
\mathbf{X}$.
As proved in Section \ref{General_solutions}, 
this motion is possible as
long as \textbf{n} and \textbf{b} are along the minor and major axes of
the elliptical section of the \textbf{B}--ellipsoid by the plane
orthogonal to \textbf{a}, respectively. 
 
Boulanger and Hayes  \cite{BoHa92} considered the propagation of
finite--amplitude \textit{homogeneous} plane waves in a deformed 
Mooney--Rivlin material and found that it was possible in any direction
of propagation along \textbf{n}. 
Here, we deal with finite amplitude \textit{inhomogeneous} plane waves
and it is the \textit{plane} of \textbf{n} and \textbf{b} 
(propagation and attenuation directions) that may be arbitrarily 
prescribed. 

Thus we construct a sinusoidal evanescent inhomogeneous plane wave as 
follows. 
First, consider any plane passing through the origin.
Let \textbf{n} and \textbf{b} be unit vectors along the respective 
major and minor axes of the elliptical section of  the 
\textbf{B}--ellipsoid by the chosen plane. 
Then the motion \eqref{expcos} is possible in the deformed 
Mooney--Rivlin material, with a linear polarization in the direction 
of $\mathbf{a} =\mathbf{b \wedge n}$, an amplitude 
exponentially attenuated by factor $k$ in the direction of \textbf{b},
and a velocity $ v \mathbf{n}$, where
$\rho v^2 = C[(\mathbf{n \cdot Bn} ) 
-(\mathbf{b \cdot Bb} )]$. 
 
In order to write explicit expressions for the directions of
\textbf{n} and \textbf{b}, and for the value of $\rho v^2 $, we use a
method developed by Boulanger and Hayes \cite[Section 5.7]{BoHa93}. 
First we prescribe a plane cutting the \textbf{B}--ellipsoid in a 
central elliptical section. 
We denote by \textbf{a} the unit vector normal to this plane. 
Then we write the Hamiltonian cyclic decomposition of the tensor 
\textbf{B} as  \cite[Section 3.4]{BoHa93} 
\begin{align} \label{Hamil} 
\mathbf{B} =  & \, \lambda_1^2 \, \mathbf{i \otimes i} 
+ \lambda_2^2 \, \mathbf{j \otimes j} 
+ \lambda_3^2 \, \mathbf{k \otimes k} \\ \nonumber 
= & \, \lambda_2^2 \, \mathbf{1} 
+ \textstyle{\frac{1}{2}} ( \lambda_1^2 - \lambda_3^2) 
(\mathbf{m^+ \otimes m^-} +\mathbf{m^- \otimes m^+}), 
\end{align} 
where the unit vectors $\mathbf{m}^{\pm}$ are in the directions of 
the `optic axes' of the  \textbf{B}--ellipsoid 
and are defined by
\begin{equation}  \label{m+-}
\sqrt{\lambda_1^2-\lambda_3^2} \, \mathbf{m}^{\pm}=  
\sqrt{\lambda_1^2-\lambda_2^2} \, \mathbf{i}  \pm  
\sqrt{\lambda_2^2-\lambda_3^2} \, \mathbf{k}.
\end{equation}  
We now seek \textbf{n} and \textbf{b}, unit vectors in the directions 
of the principal axes of the elliptical section of the 
\textbf{B}--ellipsoid  by the plane $\mathbf{a\cdot x} =0$. 
Hence, \textbf{n} and \textbf{b} are eigenvectors of the tensor 
$ \Pi \mathbf{B} \Pi$ where $\Pi$, the orthogonal projection upon 
the plane $\mathbf{a\cdot x} =0$, is defined by 
$\Pi = \mathbf{1} - \mathbf{a \otimes a}$.
With the aid of \eqref{Hamil}, and because $\Pi^2 = \Pi$, we have 
\begin{equation} \label{PiBPi} 
\Pi \mathbf{B} \Pi =  \lambda_2^2 \, \Pi 
+ \textstyle{\frac{1}{2}} ( \lambda_1^2 - \lambda_3^2) 
(\Pi \mathbf{m^+} \otimes \Pi \mathbf{m^-} 
+ \Pi \mathbf{m^-} \otimes \Pi \mathbf{m^+}).
\end{equation} 
Calling $\psi^{\pm}$ the angles between the polarization direction 
\textbf{a} and the optic axes 
$\mathbf{m^{\pm}}$ ($0 \leq \psi^{\pm} \leq \pi$), 
and noting that the 
vectors $\Pi \mathbf{m^{\pm}} / \sin \psi^{\pm}$ are of unit length, 
we find from \eqref{PiBPi} that 
$\Pi \mathbf{m^+} / \sin \psi^+ \pm \Pi \mathbf{m^-} / \sin \psi^- $ 
are eigenvectors of $ \Pi \mathbf{B} \Pi$ with eigenvalues 
$\gamma^{\pm}$ given by \cite[Section 5.7]{BoHa93}
\begin{equation} \label{gamma} 
\gamma^{\pm}=  \lambda_2^2 + \textstyle{\frac{1}{2}} 
( \lambda_1^2 - \lambda_3^2) 
(\Pi \mathbf{m^+}  \cdot \Pi \mathbf{m^-} 
\pm \sin \psi^+ \sin \psi^-). 
\end{equation} 

Recall that the phase speed $v$, which is given by equation 
\eqref{speed},
\begin{equation} 
\rho v^2 = C (\mathbf{n \cdot Bn} 
- \mathbf{b \cdot Bb}) 
= C (\mathbf{n} \cdot \Pi \mathbf{B} \Pi\mathbf{n} 
- \mathbf{b} \cdot \Pi \mathbf{B} \Pi \mathbf{b}), 
\end{equation} 
was assumed to be real. 
Therefore, \textbf{n} is the eigenvector of $ \Pi \mathbf{B} \Pi$ with
the greater eigenvalue, which is $\gamma^+$ according to 
\eqref{gamma}, and \textbf{b} is the eigenvector of 
$ \Pi \mathbf{B} \Pi$ with the lesser eigenvalue $\gamma^-$. 
Hence, we obtain the propagation and attenuation directions as  
\begin{equation}  \label{n&b} 
\begin{array}{l}  
\mathbf{n}= \Pi \mathbf{m^+} / \sin \psi^+ 
+ \Pi \mathbf{m^-} / \sin \psi^-, \quad 
\mathbf{n \cdot Bn} = \gamma^+ , \\ 
\mathbf{b}= \Pi \mathbf{m^+} / \sin \psi^+ 
- \Pi \mathbf{m^-} / \sin \psi^-, \quad 
\mathbf{b \cdot Bb} = \gamma^-. 
\end{array}  
\end{equation}

%---------------------------------------------- 
\subsection{Phase speed and pressure} 
%---------------------------------------------- 

We now write the phase speed in terms of the initial stretches and 
the angles $\psi^{\pm}$. 
Upon using \eqref{gamma} and \eqref{n&b}, we have 
\begin{equation} \label{speed/psi}
\rho v^2 = C (\lambda_1^2 - \lambda_3^2) \sin \psi^+ \sin \psi^-. 
\end{equation}
In this connection, it may be noted that for finite--amplitude
\textit{homogeneous} plane waves propagating in a homogeneously
deformed Mooney--Rivlin material, two linearly--polarized waves
exist for each direction of propagation, and the difference between
the two corresponding squared speeds is proportional to 
$(\lambda_1^2 - \lambda_3^2) \sin \phi^+ \sin \phi^-$,
where $\phi^{\pm}$  are the angles between the propagation direction
and the optic axes of the $\mathbf{B}^{-1}$--ellipsoid
(also called the `acoustic axes') \cite{BoHa92}.
 
From equation \eqref{speed/psi}, it follows that the  
minimum value of $v$ is $v_{min}$ given by 
\begin{equation} 
 v_{min}^2 = 0, 
\end{equation} 
and is attained only when $\mathbf{a}=\mathbf{m^{\pm}}$. 
In this case, 
$\mathbf{n \cdot Bn} = 
\mathbf{b \cdot Bb} = \lambda_2^2$, 
and the plane orthogonal to \textbf{a} is a plane of central 
circular section of the \textbf{B}--ellipsoid. 
In other words, when the plane of \textbf{n} and \textbf{b} is 
prescribed to be orthogonal to an optic axis of the 
\textbf{B}--ellipsoid, the superposed deformation must be static.
We treat this case in Section \ref{Superposed_static_deformation}. 
 
The maximum value for the phase speed is $v_{max}$ given by 
\begin{equation} 
\rho v_{max}^2 = C (\lambda_1^2 - \lambda_3^2) , 
\end{equation} 
which is attained when $\mathbf{a}=\mathbf{j}$. 
In that case, $\mathbf{n}= \mathbf{i}$ and $\mathbf{b} =\mathbf{k}$. 
In other words, the fastest wave occurs for propagation in the 
direction of the greatest initial stretch. 
A similar result was established by Ericksen \cite{Eric53} for 
acceleration waves in a homogeneously deformed neo--Hookean material 
and also by Boulanger and Hayes \cite{BoHa95} for finite--amplitude 
homogeneous waves in a homogeneously deformed Mooney--Rivlin material. 
 
We may also write the phase speed in terms of the polarization 
direction \textbf{a} alone. 
Because $\mathbf{n \cdot Bb} =0$ and 
det $\mathbf{B}=1$, we have 
\begin{align} \label{aB-1a} 
\mathbf{a \cdot B^{-1} a} = & \, 
(\mathbf{n \wedge b})\cdot 
\mathbf{B}^{-1} (\mathbf{n \wedge b})
= (\mathbf{n \wedge b})\cdot 
(\mathbf{Bn \wedge Bb}) \\ \nonumber 
= & \, (\mathbf{n \cdot Bn})
(\mathbf{b \cdot Bb}). 
\end{align} 
Writing the trace of \textbf{B} in the 
(\textbf{a}, \textbf{n}, \textbf{b}) basis yields 
\begin{equation} \label{trB} 
\mathrm{tr}\mathbf{B}= \mathbf{a \cdot Ba} 
+ \mathbf{n \cdot Bn} 
+ \mathbf{b \cdot Bb}. 
\end{equation} 
Combining \eqref{aB-1a} and \eqref{trB}, we see that 
$\mathbf{n \cdot Bn}$ and 
$\mathbf{b \cdot Bb}$ are the two roots of the following 
quadratic in $r$ (say), 
\begin{equation} 
r^2 + [\mathrm{tr}\mathbf{B}-\mathbf{a \cdot Ba}] r 
+ \mathbf{a \cdot B^{-1} a} =0, 
\end{equation} 
so that, using \eqref{speed}, $v$ is alternatively given by 
\begin{equation} 
\rho v^2 = C \sqrt{[\mathrm{tr} \, \mathbf{B}
-(\mathbf{a \cdot Ba})]^2 - 4(\mathbf{a \cdot B^{-1} a})}. 
\end{equation} 
 
Finally, the pressure $p^* $ is given by 
\begin{multline}  
p^* = \alpha kD e^{-\mathbf{b\cdot x}} 
[(\mathbf{a \cdot B^{-1} n}) \sin k (\mathbf{n\cdot x}-vt)  
+(\mathbf{a \cdot B^{-1} b}) \cos k (\mathbf{n\cdot x}-vt) ]\\ 
+\frac{\alpha^2  k^2}{2}D e^{-2k\mathbf{b\cdot x}} 
(\mathbf{a \cdot B^{-1} a}). 
\end{multline}

%---------------------------------------------- 
\subsection{Energy propagation} 
%---------------------------------------------- 
 
Here we consider the energy carried by the wave \eqref{expcos}. 
First we compute the total energy density, which is the sum of the 
kinetic and stored--energy densities, and then the energy flux vector. 
Our aim is to find a relationship between these two quantities, or 
more relevantly (as the frequencies of the sinusoidal vibrations may 
be very high), between the temporal mean values of these quantities.  
To this effect, we introduce the following notation to designate 
temporal mean values: if $D(\mathbf{x},t)$ is a periodic field 
quantity with frequency $\omega$, then its mean value is $\check{D}$, 
defined by 
\begin{equation} 
\check{D}= \frac{\omega}{2\pi} \int_{0}^{\frac{\omega}{2\pi}} 
D(\mathbf{x},t)dt. 
\end{equation} 
 
We begin  with the kinetic energy density per unit 
volume $\overline{K}$ given  by equation \eqref{kinetic_energy}.
Using \eqref{expcos}, its mean value $\check{\overline{K}}$ is found 
to be
\begin{equation} \label{K} 
\check{\overline{K}} 
= \frac{\alpha^2 k^2}{4} e^{-2k\mathbf{b\cdot x}} \rho v^2 
= \frac{\alpha^2 k^2}{4} e^{-2k\mathbf{b\cdot x}} 
C [(\mathbf{n \cdot Bn})  
-(\mathbf{b\cdot Bb} )]. 
\end{equation} 
 
The stored--energy density $\overline{W}$  per unit volume  
associated with the wave is given by \eqref{stored_energy}, 
where $\overline{I}$ and $\overline{II}$ are given, 
for the motion \eqref{expcos}, by 
\begin{equation}  
\begin{array}{l}
\begin{split} 
\overline{I}  =  I 
-& 2  \alpha k e^{-k \xi} [ \cos k(\mathbf{n\cdot x}-vt) 
(\mathbf{a \cdot Bb} ) 
+  \sin k(\mathbf{n\cdot x}-vt) (\mathbf{a\cdot Bn}) ] \\ 
 + & \alpha^2 k^2 e^{-2k \xi} [ \cos^2 k(\mathbf{n\cdot x}-vt)
(\mathbf{b \cdot Bb}) 
+ \sin^2 k(\mathbf{n\cdot x}-vt) (\mathbf{n \cdot Bn})], 
\end{split} \vspace{2mm} \\ 
\begin{split} 
\overline{II}  =  II 
+ 2 \alpha k e^{-k \xi} [ & \cos k(\mathbf{n\cdot x}-vt) 
(\mathbf{a \cdot B^{-1} b} ) \\
 & + \sin  k(\mathbf{n\cdot x}-vt) 
(\mathbf{a \cdot B^{-1} n})] 
 +\alpha^2 k^2 e^{-2k \xi} (\mathbf{a \cdot B^{-1} a}). 
\end{split} 
\end{array}
\end{equation} 
Note that because the material is incompressible, the stored--energy 
density $\overline{W}$ is the same whether it is measured 
in the reference configuration, in the  state of static homogeneous 
deformation, or in the current configuration. 
The mean value of this quantity is $\check{\overline{W}}$, given by 
\begin{equation} \label{W} 
\check{\overline{W}} = \frac{\alpha^2 k^2}{4}e^{-2k\mathbf{b\cdot x}} 
[C(\mathbf{n \cdot Bn}
+\mathbf{b \cdot Bb} )  
+2D(\mathbf{a \cdot B^{-1} a} )]. 
\end{equation} 
 
We can now compute the total energy density $\overline{E}$, which is 
by definition the sum of the kinetic and stored--energy densities: 
$\overline{E} = \overline{K} + \overline{W}$. 
Using \eqref{K} and \eqref{W}, we write directly the mean value 
$\check{\overline{E}}$ of the total energy density  $\overline{E}$ as 
\begin{equation} \label{E} 
\check{\overline{E}} = \frac{\alpha^2 k^2}{2}e^{-2k\mathbf{b\cdot x}} 
[C(\mathbf{n \cdot Bn})  
+D(\mathbf{a \cdot B^{-1} a} )]. 
\end{equation} 
 
Now we turn our attention to the energy flux vector,
defined in equation \eqref{energy_flux}, and find 
that here, \textbf{R} is given by 
\begin{multline} 
\mathbf{R} =  
-\alpha vk e^{-k \mathbf{b\cdot x}} \sin k (\mathbf{n\cdot x}-vt)  
 \{ \mathbf{Ta} 
- \frac{\alpha^2 k^2}{2} D e^{-2k\mathbf{b\cdot x}}  
(\mathbf{a \cdot B^{-1} a})\mathbf{a} \\ 
 - \alpha  k 
[C \mathbf{Bb}
+ D(\mathbf{a \cdot B^{-1} a})\mathbf{n}] 
e^{-k\mathbf{b\cdot x}} \cos k(\mathbf{n\cdot x}-vt) \\ 
  - \alpha  k 
[C \mathbf{Bn}+ 
D(\mathbf{a \cdot B^{-1} a})\mathbf{n}] 
e^{-k\mathbf{b\cdot x}} \sin k (\mathbf{n\cdot x}-vt) \}. 
\end{multline} 
The mean value of \textbf{R} is $\check{\mathbf{R}}$, given by 
\begin{equation}\label{R} 
\check{\mathbf{R}} = 
\frac{\alpha^2 k^2}{2}v e^{-2k\mathbf{b\cdot x}}  
[C\mathbf{Bn}
+D(\mathbf{a \cdot B^{-1}a})\mathbf{n}]. 
\end{equation} 
 
Hence, using equation \eqref{conj}, we see that 
\begin{equation} \label{R.b} 
\check{\mathbf{R}}\cdot \mathbf{b} =0, 
\end{equation} 
which means that the direction of the mean energy flux vector  
is parallel to the planes of constant amplitude. 
We also have, using \eqref{E},
 \begin{equation}  \label{R.n} 
\check{\mathbf{R}}\cdot (v^{-1} \mathbf{n})= \check{\overline{E}}, 
\end{equation} 
which means that the component of the mean energy flux vector
in the direction of propagation is
 equal to the phase speed times the total energy density. 
 
These results may be compared with results established
previously. 

First, Schouten \cite[Section VII.7]{Scho51} introduced the notion of
an energy flux vector for the propagation of elastic waves in
anisotropic elastic media. 
For small homogeneous displacements (that is for displacements of the 
form $ \mathbf{a}\cos (kx -vt)$, where the real vector \textbf{a} is 
of infinitesimal magnitude), he proved that the temporal mean values 
of the energy flux vector and energy densities are related through 
the same equation as \eqref{R.n}.  
  
Then Synge \cite{Syng56} looked for solutions to the equations of 
motion in an anisotropic medium in the form 
$ \{ \mathbf{A}e^{i\omega(\mathbf{S\cdot x}-t)}  + \mathrm{c.c.} \}$ 
where $\mathbf{A}=\mathbf{A}^+ + i\mathbf{A}^-$ and
$\mathbf{S}=\mathbf{S}^+ + i\mathbf{S}^-$ are complex vectors, 
 $\omega$ is the real frequency and `c.c.' denotes the
complex conjugate. 
He proved that 
$\check{\mathbf{R}} \cdot \mathbf{S}^- = 0$ 
and $\check{\mathbf{R}} \cdot \mathbf{S}^+  \geq 0$, 
where $\check{\mathbf{R}}$ is the time-averaged energy flux vector. 
 
Then Hayes \cite{Haye75} showed  for \textit{any} linear 
conservative system,  using the same notation as above, that
$\check{\mathbf{R}} \cdot \mathbf{S}^- = 0$ 
and $\check{\mathbf{R}} \cdot \mathbf{S}^+= \check{E}$, 
where $\check{E}$ is the mean energy density carried by the wave. 
In that paper, no assumption is made as to whether or not the medium 
is anisotropic or subject to an internal constraint such as 
incompressibility or inextensibility.
 
Later, Chadwick, Whitworth and Borejko \cite{ChWB85} and Borejko
\cite{Bore87} described the dynamics of small amplitude plane waves
superposed upon a large homogeneous deformation of a constrained
material, be they homogeneous \cite{ChWB85} or inhomogeneous
\cite{Bore87}. 
In order to remain consistent with results of linear
elastodynamics, these authors (following Lighthill \cite{Ligh78})
separated energy quantities into `interaction' and `incremental'
parts. The former terms refer to the interconnection between the
primary static stretch and the superposed waves, whereas the latter
refer to the waves only, and would not disappear in the absence of
prestress. This distinction made, they proved that
$\check{\mathbf{R}}^{incr} \cdot \mathbf{S}^- = 0$ 
and
$\check{\mathbf{R}}^{incr} \cdot \mathbf{S}^+ = \check{E}^{incr}$,
where the superscript `incr' is short for `incremental'.
 
These results are all similar to those given by \eqref{R.b} and 
\eqref{R.n} because, in our context, the displacement 
$\overline{\mathbf{x}}-\mathbf{x}$ may also be written in the form 
$ \{ \mathbf{A}e^{i\omega(\mathbf{S\cdot x}-t)} + \mathrm{c.c.}\}$,  
with $\mathbf{A}= \alpha \mathbf{a}/2$, $\omega = vk$ and 
$\mathbf{S}=v^{-1} (\mathbf{n}+i\mathbf{b})$. 
With this notation, \eqref{R.b} and \eqref{R.n} are rewritten as 
$\check{\mathbf{R}}  \cdot \mathbf{S}^- = 0$ and 
$\check{\mathbf{R}} \cdot \mathbf{S}^+ = \check{\overline{E}}$. 
 
However, the above mentioned studies are situated within the 
framework of \textit{linearized} theory and it is remarkable that 
results such as equations \eqref{R.b} and \eqref{R.n} may be found in 
the non-linear case of a \textit{finite}--amplitude wave propagating 
in a finitely deformed Mooney--Rivlin material. 
Note that in the same context, 
Boulanger and Hayes \cite{BoHa95} found  
equation \eqref{R.n} for finite--amplitude \textit{homogeneous}
plane waves.

%+++++++++++++++++++++++++++++++++++++++++++++++++++++++++++++++++++ 
\section{Superposed static deformation} 
\label{Superposed_static_deformation} 
%+++++++++++++++++++++++++++++++++++++++++++++++++++++++++++++++++++ 
 
An interesting feature of inhomogeneous plane motions 
is that for certain 
orientations of the planes of constant phase and constant amplitude, 
the speed $v$ may be equal to zero \cite[Section 6.3]{BoHa93}.
In this situation, no perturbation propagates, 
but the static deformation 
$f(\mathbf{b\cdot x})g(\mathbf{n\cdot x}) \mathbf{a}$ 
may nevertheless be superposed 
upon the primary homogeneous one.
This possibility provides new exact solutions to the equations of 
equilibrium in a Mooney--Rivlin material.

Here, we consider in turn the case of special
and  non--special  principal inhomogeneous deformations. 

%----------------------------------------------------------------- 
\subsection{Superposed displacements along principal axes of the 
\textbf{B}--ellip\-soid} 
%----------------------------------------------------------------- 
 
For special principal motions, 
we saw in Section \ref{Special_principal_motions} that the quantity
$v$ can be prescribed within the interval 
$0 \leq v^2 < (C \lambda_1^2 + D \lambda_2^{-2})/\rho$.
When we choose $v$ to be zero, we obtain from \eqref{SpecialMtn}
a possible static deformation of a Mooney--Rivlin material. 
It is written as
\begin{equation} 
\overline{x}= \lambda_1 X, \quad
\overline{y}= \lambda_2 Y +f(\lambda_3 Z) g(\lambda_1 X), \quad 
\overline{z}= \lambda_3 Z , 
\end{equation}
where either
\begin{equation} \label{static1}
\left.
\begin{array}{l}
f(\lambda_3 Z)= 
a_1 \exp k \sqrt{ \frac{C \lambda_1^2 + D \lambda_2^{-2}}
{C \lambda_3^2 + D \lambda_2^{-2}} } \lambda_3 Z
 +a_2 \exp -k \sqrt{\frac{C \lambda_1^2 + D \lambda_2^{-2}}
{C \lambda_3^2 + D \lambda_2^{-2}}} \lambda_3 Z, 
\vspace{2mm}\\
g(\lambda_1 X)=
d_1 \cos k (\lambda_1 X) + d_2 \sin k (\lambda_1 X),
\end{array}
\right\}
\end{equation} 
or 
\begin{equation} \label{static2}
\left.
\begin{array}{l}
f(\lambda_3 Z)=
a_1 \cos k \sqrt{ \frac{C \lambda_1^2 + D \lambda_2^{-2}}
{C \lambda_3^2 + D \lambda_2^{-2}} } \lambda_3 Z
 +a_2 \sin k  \sqrt{ \frac{C \lambda_1^2 + D \lambda_2^{-2}}
{C \lambda_3^2 + D \lambda_2^{-2}} } \lambda_3 Z, 
\vspace{2mm}\\
g(\lambda_1 X)= 
 d_1 \exp k (\lambda_1 X) 
+ d_2 \exp - k(\lambda_1 X).
\end{array}
\right\}
\end{equation} 
Here, $a_1, a_2, d_1, d_2$ and $k$ are arbitrary constants.

The components of the Cauchy stress corresponding to these
deformations are as follows:
\begin{equation}
\begin{array}{l} 
\overline{\mathbf{T}}_{11}= -\overline{p} +C \lambda_1^2 
-D [\lambda_1^{-2} + \lambda_2^{-2} ( fg')^2] ,      \\ 
\overline{\mathbf{T}}_{12}= 
(C \lambda_1^2 + D \lambda_2^{-2}) fg' ,            \\ 
\overline{\mathbf{T}}_{13}= -D \lambda_2^{-2} ff'gg' ,      \\ 
\overline{\mathbf{T}}_{22}= -\overline{p}
+ C[ \lambda_1^2 ( fg')^2 +\lambda_2^2
+\lambda_3^2 ( f'g)^2] -D \lambda_2^{-2},\\ 
\overline{\mathbf{T}}_{23}= 
(C \lambda_3^2 + D \lambda_2^{-2}) f'g ,               \\ 
\overline{\mathbf{T}}_{33}=-\overline{p} +C \lambda_3^2 
-D [\lambda_3^{-2} + \lambda_2^{-2} ( f'g)^2],   
\end{array} 
\end{equation} 
where the pressure $\overline{p} $ is given either by
\begin{equation}
\overline{p}=
\frac{D k^2}{2} \lambda_2^{-2}
[(d_1^2 + d_2^2) f^2 - 
4 \left( \frac{C \lambda_1^2 + D \lambda_2^{-2}}
{C \lambda_3^2 + D \lambda_2^{-2}} \right) a_1 a_2 g^2],
\end{equation} 
when $f$ and $g$ are given by \eqref{static1}, or by
\begin{equation}
\overline{p}=
-\frac{D k^2}{2} \lambda_2^{-2}
[4d_1^2 d_2^2 f^2 -
\left( \frac{C \lambda_1^2 + D \lambda_2^{-2}}
{C \lambda_3^2 + D \lambda_2^{-2}} \right) ( a_1^2 + a_2^2) g^2],
\end{equation} 
when $f$ and $g$ are given by \eqref{static2}.

%-------------------------------------------------------------------- 
\subsection{Superposed displacements in the planes of central
circular section of the \textbf{B}--ellipsoid} 
%-------------------------------------------------------------------- 

From Section \ref{General_solutions}, 
we see that for certain choices of the
orthogonal unit vectors \textbf{n} and \textbf{b}, the quantity  
$v$ is equal to zero. 
This is the case when the plane of the conjugate vectors \textbf{n} 
and \textbf{b} is a plane of central circular section of the 
\textbf{B}--ellipsoid because then, 
$\mathbf{n \cdot Bn}=\mathbf{b \cdot Bb}
=\lambda_2^2$ (see \cite[Section 5.7.1]{BoHa93} for instance) and
then, from equation \eqref{speed}, $v=0$. 
The vector \textbf{a} is one of the unit vectors $\mathbf{m}^{\pm}$,
which are in the directions of the optic axes of the 
\textbf{B}--ellipsoid and are defined by \eqref{m+-}.
Also, (\textbf{n}, \textbf{b}) is any pair of orthogonal unit vectors 
in a plane of central circular section 
$\mathbf{a\cdot x}=\mathbf{m^{\pm} \cdot x}=0$. 

To provide an illustrative example, we choose 
$\mathbf{n}=\mathbf{m}^{\pm} \wedge
\mathbf{j}$, $\mathbf{b}= \mathbf{j}$, and 
$f(\mathbf{b\cdot x})g(\mathbf{n\cdot x}) = 
\alpha e^{-k\mathbf{b\cdot x}} \cos k\mathbf{n\cdot x} $, 
where $\alpha$ is a constant.
Hence, two possible finite static deformations of a 
Mooney--Rivlin material are 
\begin{equation} \label{static} 
\left. 
\begin{array}{l} 
x= \lambda_1 X + \alpha 
\sqrt{\frac{\lambda_1^2-\lambda_2^2}{\lambda_1^2-\lambda_3^2}} 
e^{-k  \lambda_2 Y} 
\cos k \big{(} \lambda_1 
\sqrt{\frac{\lambda_2^2-\lambda_3^2}{\lambda_1^2-\lambda_3^2}}X 
\mp \lambda_3 \sqrt{\frac{\lambda_1^2-\lambda_2^2}
{\lambda_1^2-\lambda_3^2}}Z \big{)}, \\ 
y= \lambda_2 Y, \\ 
z= \lambda_3 Z \pm \alpha 
\sqrt{\frac{\lambda_2^2-\lambda_3^2}{\lambda_1^2-\lambda_3^2}} 
e^{-k  \lambda_2 Y} 
\cos k \big{(}\lambda_1 
\sqrt{\frac{\lambda_2^2-\lambda_3^2}{\lambda_1^2-\lambda_3^2}}X 
\mp \lambda_3 \sqrt{\frac{\lambda_1^2-\lambda_2^2}
{\lambda_1^2-\lambda_3^2}}Z \big{)}. 
\end{array} 
\right\} 
\end{equation}

The stress tensor $\overline{\mathbf{T}}$ components for these 
deformations can be written in the basis 
$(\mathbf{n}, \mathbf{a}, \mathbf{b})= 
(\mathbf{m}^{\pm} \wedge \mathbf{j}, \mathbf{m}^{\pm}, \mathbf{j})$ 
as 
\begin{equation} 
\begin{array}{l} 
\mathbf{n \cdot \overline{T}  n}
=\mathbf{n \cdot Tn} 
-p^* +\alpha^2 k^2 D \lambda_2^4 e^{-2k\zeta} \cos^2 k\eta ,\\ 
\mathbf{n \cdot \overline{T}  a}
=\mathbf{n\cdot Ta} 
-\alpha k (C+ D \lambda_2^2)\lambda_2^2 e^{-k\zeta} \sin k\eta ,
\vspace{2mm}\\ 
\begin{split} 
\mathbf{n \cdot \overline{T}  b}=  
\mp \alpha k D  \lambda_2^2  
\sqrt{(\lambda_1^2 -\lambda_2^2) (\lambda_2^2 -\lambda_3^2)} 
& e^{-k \zeta} \cos k\eta \\ 
 - & \alpha^2 k^2 D \lambda_2^4 e^{-2k \zeta} \sin k\eta \cos k\eta,  
\end{split}
\vspace{2mm}\\ 
\begin{split}
\mathbf{a \cdot \overline{T}  a} 
=\mathbf{a\cdot Ta} -p^* \pm 2\alpha k C 
\sqrt{(\lambda_1^2 -\lambda_2^2)(\lambda_2^2 -\lambda_3^2)} 
& e^{-k \zeta} \sin k\eta  \\
+ & \alpha^2 k^2 C \lambda_2^2 e^{-2k\zeta}, 
\end{split} \\ 
\mathbf{a \cdot \overline{T}b}= 
-\alpha k (C+ D \lambda_2^2)\lambda_2^2 e^{-k\zeta} \cos k\eta, 
\vspace{2mm}\\ 
\begin{split} 
\mathbf{b \cdot \overline{T}  b}
=\mathbf{b\cdot Tb} 
-p^*  \pm 2\alpha k D \lambda_2^2 
& \sqrt{(\lambda_1^2 -\lambda_2^2) (\lambda_2^2 -  \lambda_3^2)} 
e^{-k \zeta} \sin k\eta \\ 
  &  \phantom{\sqrt{(\lambda_1^2 -\lambda_2^2) }}
+ \alpha^2 k^2 D \lambda_2^4 e^{-2k\zeta} \sin^2 k\eta, 
\end{split} 
\end{array} 
\end{equation} 
where \textbf{T} is the constant Cauchy stress tensor  
defined by equation \eqref{T} and the pressure $p^*$ is given by 
\begin{equation} 
p^*=\pm \alpha k D  \lambda_2^2  
\sqrt{(\lambda_1^2 -\lambda_2^2)(\lambda_2^2 -\lambda_3^2)} 
e^{-k \zeta} \sin k\eta 
+\frac{\alpha^2 k^2}{2} D \lambda_2^4 e^{-2k \zeta}. 
\end{equation} 
 
%++++++++++++++++++++++++++++++++++++++++++++++++++++++++++++++++ 
\section{Concluding remarks} 
%++++++++++++++++++++++++++++++++++++++++++++++++++++++++++++++++ 
 
New solutions to the equations of motion and equilibrium in a 
deformed Mooney--Rivlin material have been obtained. 
In the case of a finite exponential sinusoidal wave, the propagation 
of energy was examined and results formerly established within the 
framework of linearized elasticity were recovered, although  
the case of a finite motion is non-linear. 
 
The waves considered here are \textit{linearly-polarized} and it 
ought to be noted that this is the only possibility of having  
finite--amplitude \textit{inhomogeneous} plane waves propagating in a
material (deformed or not) subjected to the constraint of
incompressibility. This is shown elsewhere \cite{Dest99}.  

Also, the planes of constant phase (orthogonal to \textbf{n})
were assumed to be at right angles with 
 the planes of constant amplitude (orthogonal to \textbf{b}).
We show in Appendix B that for finite--amplitude
inhomogeneous plane waves of complex exponential type,
this condition must be satisfied.

%++++++++++++++++++++++++++++++++++++++++++++++++++++++
% bibliography
%++++++++++++++++++++++++++++++++++++++++++++++++++++++

%++++++++++++++++++++++++++++++++++++++++++++++++++++++++++++++++ 
\appendix 
 \renewcommand{\thesection}{\Alph{section}}
%++++++++++++++++++++++++++++++++++++++++++++++++++++++++++++++++

%----------------------------------------------------------------
\section*{Appendix A} 
\textit{Solutions for a system of two linear 
differential equations involving two functions of independent 
variables.} 
\setcounter{section}{1} 
%---------------------------------------------------------------

Here, our aim is to find the real 
functions $f(\zeta)$ and $g(\eta -vt)$ 
satisfying the equations \eqref{2ndOrder} and \eqref{compat}, viz
\begin{equation} \label{syst} 
\left.
\begin{array}{l}  
\rho v_b^2 \, g f''  
+ 2  C(\mathbf{n\cdot Bb}) \, g'f' 
+ \rho (v_n^2 - v^2) \, g''f = 0, \\ 
\begin{split} 
f'''g & (\mathbf{a \cdot B^{-1} n})  
-f''g'(\mathbf{a \cdot B^{-1} b}) 
+(f'f''-ff''')gg'(\mathbf{a \cdot B^{-1} a}) 
\\ & 
=fg''' (\mathbf{a \cdot B^{-1} b}) 
-f'g''(\mathbf{a \cdot B^{-1} n}) 
+(g'g''-gg''')ff'(\mathbf{a \cdot B^{-1} a}), 
\end{split} 
\end{array} 
\right\}  
\end{equation} 
where 
\begin{equation} \label{Vb,Vn}
\rho v_b^2 = C (\mathbf{b\cdot Bb})  
+D(\mathbf{a \cdot B^{-1} a}),   \quad
\rho v_n^2 = C(\mathbf{n\cdot Bn}) 
+D(\mathbf{a \cdot B^{-1} a}),
\end{equation} 
and \textbf{B} is the left Cauchy-Green strain tensor  associated 
with an arbitrary triaxial stretch. 
Also, $f$ and $g$ are required to be such that 
$f(\mathbf{b\cdot x})g(\mathbf{n\cdot x}-vt)\mathbf{a}$ (where  
(\textbf{a}, \textbf{n}, \textbf{b}) is an orthonormal triad) is 
an \textit{inhomogeneous} deformation. 
 
A referee very kindly suggested that the use of a technique by 
Birkhoff \cite{Birk50} would lead to a simplier derivation of the 
results. Accordingly, this approach is adopted here.

Assuming $\mathbf{n\cdot Bb} \neq 0$, we can, 
following \mbox{Birkhoff} \cite[p.~137]{Birk50}, express 
equation $\eqref{syst}_1$ in the separated form
\begin{equation} \label{F.G=0}
\sum_{j=1}^{3} F_j(\zeta) G_j(\chi) = 0, \quad \chi = \eta-vt,
\end{equation}
where $F_1= f''$, $F_2= f'$,  $F_3= f$,  and
$G_1= \rho v_b^2 \, g$, $G_2= 2  C(\mathbf{n\cdot Bb}) \, g'$,
  $G_3=  \rho (v_n^2 - v^2) \, g''$.

The condition \eqref{F.G=0} is equivalent to the requirement that
the vectors  $\mathbf{F}= (F_1, F_2, F_3)$ and 
$\mathbf{G}= (G_1, G_2, G_3)$ be confined to orthogonal subspaces.
Calling $\mathrm{dim} \, \mathbf{F}$ and $\mathrm{dim} \,\mathbf{G}$
the respective dimensions of the subspaces spanned by the $\mathbf{F}$
and $\mathbf{G}$, we may have three possible cases:

Case (i): $\mathrm{dim} \, \mathbf{F} =\mathrm{dim} \,\mathbf{G}=1$;

Case (ii): $\mathrm{dim} \, \mathbf{F} =1$, 
$\mathrm{dim} \,\mathbf{G}=2$;

Case (iii): $\mathrm{dim} \, \mathbf{F} =2$, 
$\mathrm{dim} \,\mathbf{G}=1$.

We treat these cases in turn.

\underline{Case (i)}: 
The dimension of the  subspace spanned by the $\mathbf{F}$ is 1
when $F_1$, $F_2$, $ F_3$, or equivalently
$f$, $f'$, $f''$  are proportional.
This is possible only when $f(\zeta)= A e^{ \delta \zeta}$, 
where $A$ and $\delta$  are real constants. 
Similarly, $\mathrm{dim} \,\mathbf{G}=1$ yields
$g(\eta-vt)= B e^{\epsilon \chi}$ for some real scalars 
 $B$ and $\epsilon$.

However, we now have  
$fg\mathbf{a} =AB  
\exp [(\epsilon \mathbf{n}
+\delta\mathbf{b})\cdot \mathbf{x}-\epsilon vt)]
\mathbf{a}$. 
Hence, \mbox{Case (i)} corresponds to a homogeneous motion and must be
discarded in our context.

\underline{Case (ii)}:
The condition $\mathrm{dim} \, \mathbf{F} =1$ yields
\begin{equation} \label{f(ii)}
f(\zeta)= A e^{ \delta \zeta},
\end{equation} 
where $A$ and $\delta$  are real constants.
Then equation $\eqref{syst}_1$ reduces to 
\begin{equation}  \label{g(ii)}
\rho (v_n^2 - v^2) \, g''  
+ 2  C(\mathbf{n\cdot Bb})  \delta \, g' 
+\rho v_b^2  \delta^2 \, g = 0,
\end{equation} 
This is a linear homogeneous second order differential equation
for $g$, with the following characteristic equation in $r$ (say),
\begin{equation}  \label{r(ii)}
\rho (v_n^2 - v^2) \, r^2
+ 2  C(\mathbf{n\cdot Bb})  \delta \, r
+\rho v_b^2  \delta^2 = 0.
\end{equation}

If \eqref{r(ii)} has two distinct roots $r_1, r_2$ (say),
then $g$ is of the form
\begin{equation} \label{g(ii)distinct}
g(\chi)= B e^{ r_1 \chi} + C e^{ r_2 \chi},
\end{equation} 
where $B$ and $C$ are constants ($(B, C) \neq (0,0)$).

Upon using \eqref{f(ii)} and  \eqref{g(ii)distinct},  
equation $\eqref{syst}_2$ yields a linear combination 
for the independent functions 
$e^{ r_1 \chi}$, $e^{ r_2 \chi}$, $e^{2 r_1 \chi}$, $e^{2 r_2 \chi}$, 
and $e^{ (r_1+r_2) \chi}$. 
Nullity for the coefficient of  $e^{ (r_1+r_2) \chi}$ yields
$r_1 = -r_2$.
However, in that case we have $g'' - r_1^2 g=0$, which, 
together with \eqref{g(ii)}, implies that 
$\mathrm{dim} \,\mathbf{G}=1$.   

If \eqref{r(ii)} has a double  root $r$ (say),
then $g$ is of the form
\begin{equation} \label{g(ii)double}
g(\chi)= (B + C \chi) e^{ r \chi},
\end{equation} 
where $B$ and $C$ are constants ($(B, C) \neq (0,0)$).

Substituting \eqref{f(ii)} and  \eqref{g(ii)double}  into
 $\eqref{syst}_2$
yields a linear combination for the independent functions 
$e^{ r \chi}$, $\chi e^{ r \chi}$, $e^{2 r \chi}$, 
and  $\chi e^{2 r \chi}$. 
Nullity for the coefficient of  $e^{2 r \chi}$ yields
$C^2 r = 0$, which is not possible in our context.

We conclude that when $\mathbf{n\cdot Bb} \neq 0$, no $f$ and $g$ can
be found such that 
$fg \, \textbf{a}$  is 
an inhomogeneous deformation. 
 We must therefore assume that  
\begin{equation}  \label{bBn=0}
\mathbf{b \cdot Bn}=0, 
\end{equation} 
which means that the orthogonal unit vectors \textbf{n} and 
\textbf{b} are conjugate with respect to the \textbf{B}--ellipsoid. 
 
Then, equation \eqref{syst}$_1$ reduces to
\begin{equation} \label{f/g}
\frac{ v_b^2}{v_n^2 - v^2} \frac{f''(\zeta)}{f(\zeta)} 
=-\frac{g''(\eta -vt)}{g(\eta -vt)}= \mathrm{const.}= \pm k^2 
 \quad \mathrm{(say)}. 
\end{equation} 

Clearly, from \eqref{Vb,Vn} and the strong ellipticity condition 
\eqref{ellip}$_1$, we have $ v_b^2 >0$ and  $ v_n^2 >0$. 
Now, if  $ v^2 > v_n^2$, then by \eqref{f/g}, $f''/f$ and 
$g''/g$ are constants of the same sign. 
In this case, $f$ and $g$ would be of the same type (either both
 hyperbolic or both sinusoidal functions), and  again this would lead 
to a homogeneous motion. 
We assume therefore that $ 0 \leq v^2 < v_n^2$. 

Upon using \eqref{f/g},  the remaining equation to be satisfied, 
$\eqref{syst}_2$, reduces to
\begin{equation} \label{cond}
( v_n^2 -v_b^2 - v^2)
\left[ f' g (\mathbf{a \cdot B^{-1} n})
-f g' (\mathbf{a \cdot B^{-1} b}) \right] =0.
\end{equation}

Hence, there are two cases: (i) $ v^2 \neq v_n^2 - v_b^2$ and 
(ii)  $ v^2 = v_n^2 - v_b^2$.

\medskip
\underline{Case (i)}: if $ v^2 \neq v_n^2 - v_b^2$, 
then by \eqref{cond}, we have
\begin{equation} \label{mu}
\frac{f'(\zeta)}{f(\zeta)} (\mathbf{a \cdot B^{-1} n})
=  \frac{g'(\eta -vt)}{g(\eta -vt)}(\mathbf{a \cdot B^{-1} b}) 
= \mathrm{const.} = \mu \quad (\mathrm{say}). 
\end{equation} 

 Assuming $\mu \neq 0$,  \eqref{mu}
yields $(\mathbf{a \cdot B^{-1} n}) \neq 0$,
$(\mathbf{a \cdot B^{-1} b}) \neq 0$ and also  
$(\mathbf{a \cdot B^{-1} n})^2 (f''/f)
= (\mathbf{a \cdot B^{-1} b})^2 (g''/g) =
\mu^2$. 
However, as noted above, $f''/f$ and $g''/g$ must be
constants of opposite signs for an inhomogeneous motion. 

Hence, $\mu = 0$. 
Then, by \eqref{mu}, $(\mathbf{a \cdot B^{-1} n}) 
=(\mathbf{a \cdot B^{-1} b}) = 0$, which, together with 
\eqref{bBn=0}, implies that \textbf{n}, \textbf{a} and \textbf{b} are 
along principal directions. 
In this case, the solutions are either exponential in space and 
sinusoidal in time,
\begin{equation} 
\left.
\begin{array}{l}
f(\zeta)=a_1 \exp k \sqrt{ \frac{v_n^2 - v^2}{v_b^2}}\zeta
 +a_2 \exp -k \sqrt{ \frac{v_n^2 - v^2}{v_b^2}}\zeta, \\
g(\eta-vt)= d_1 \cos k (\eta-vt) + d_2 \sin k(\eta -vt),
\end{array}
\right\}
\end{equation} 
or sinusoidal in space and exponential in time,
\begin{equation} 
\left.
\begin{array}{l}
f(\zeta)=a_1 \cos k \sqrt{ \frac{v_n^2 - v^2}{v_b^2}}\zeta
 +a_2 \sin k \sqrt{ \frac{v_n^2 - v^2}{v_b^2}}\zeta, \\
g(\eta-vt)= d_1 \exp k (\eta-vt) + d_2 \exp - k(\eta -vt).
\end{array}
\right\}
\end{equation} 
Here, $a_1, a_2, d_1, d_2$ are constants, and $k$ and $v$ are
arbitrary $(0 \leq v^2 <v_n^2)$.

These solutions are valid \textit{only} when \textbf{n}, \textbf{a} 
and \textbf{b} are along the principal axes of the primary static 
deformation.

\medskip
\underline{Case (ii)}: if $ v^2 = v_n^2 - v_b^2$, 
then the solutions are either 
exponential in space and sinusoidal in time,
\begin{equation} 
\left.
\begin{array}{l}
f(\zeta)=a_1 e^{ k \zeta} +a_2 e^{-k \zeta}, \\
g(\eta-vt)= d_1 \cos k (\eta-vt) + d_2 \sin k(\eta -vt),
\end{array}
\right\}
\end{equation} 
or sinusoidal in space and exponential in time,
\begin{equation} 
\left.
\begin{array}{l}
f(\zeta)=a_1 \cos k \zeta +a_2 \sin k \zeta, \\
g(\eta-vt)= d_1 e^{k (\eta-vt)} + d_2 e^{ - k(\eta -vt)}.
\end{array}
\right\}
\end{equation} 
Here, $a_1, a_2, d_1, d_2$ are constants, $k$ is
arbitrary  and $v$ is given by
\begin{equation}  
\rho v^2 = C[(\mathbf{n \cdot Bn} )  
-(\mathbf{b \cdot Bb} )].
\end{equation} 

These solutions are valid for \textit{any} orientation  of the
plane of \textbf{n} and  \textbf{b}.

%++++++++++++++++++++++++++++++++++++++++++++++++++++++++++
\section*{Appendix B}
\textit{Finite--amplitude inhomogeneous plane waves 
of complex exponential type
in deformed Mooney--Rivlin materials.} 
\setcounter{section}{2}\setcounter{equation}{0}
%+++++++++++++++++++++++++++++++++++++++++++++++++++++++++++

Here, we prove that finite--amplitude inhomogeneous 
plane waves of complex exponential type may propagate in a 
homogeneously deformed Mooney--Rivlin material only when 
the planes of constant phase are at right angles with the
planes of constant amplitude.
\vspace{12pt}

The  material is first subjected to  a pure homogeneous 
static deformation,
with  corresponding deformation gradient $\mathbb{F}$ and left 
Cauchy--Green strain tensor $\mathbb{B}$  given by
\begin{equation} 
\mathbb{F}= \mathrm{diag} \, (\lambda_1, \lambda_2, \lambda_3)
, \quad
\mathbb{B}= \mathrm{diag} \, (\lambda_1^2, \lambda_2^2, 
\lambda_3^2),
\quad \mathrm{ with} \quad J=\mathrm{det} \, \mathbb{F} =
\lambda_{1}\lambda_{2}\lambda_{3}.
\end{equation}

Then  a linearly--polarized inhomogeneous plane wave of 
finite amplitude is superposed upon the large static deformation. 
The motion is given by
\begin{equation} \label{xBar(B)}
\overline{\mathbf{x}} = \mathbf{x}+ \beta
\{ e^{i \omega(\mathbf{S\cdot x}-t)} +\mathrm{c.c.} \}\mathbf{a}
= \mathbf{x}+ 2\beta  e^{- \omega \mathbf{S^- \cdot x}}
\cos \omega(\mathbf{S^+ \cdot x}-t) \mathbf{a}.
\end{equation}
Here, $\beta$ is a finite real scalar, 
$\mathbf{S}= \mathbf{S}^+ +i \mathbf{S}^-$ is a complex vector, 
$\omega$ is the real frequency and `c.c.' denotes the complex 
conjugate. 

The deformation gradient $\overline{\mathbb{F}}$ associated with 
the motion \eqref{xBar(B)} is given by 
\begin{equation}
\overline{\mathbb{F}} 
=\partial \overline{\mathbf{x}} / \partial \mathbf{X}
=\check{\mathbb{F}} \mathbb{F},
\quad \mathrm{where} \quad \check{\mathbb{F}}=
\mathbf{1}
+ \beta \omega \mathbf{a} \otimes  
\{ i  e^{i \omega(\mathbf{S\cdot x}-t)}\mathbf{S} 
+\mathrm{c.c.} \}.
\end{equation}
The  left Cauchy--Green tensor is $\overline{\mathbb{B}}$ given by
$\overline{\mathbb{B}} =
\overline{\mathbb{F}}\, \overline{\mathbb{F}}^{\mathrm{T}} =
\check{\mathbb{F}}\mathbb{B}\check{\mathbb{F}}^{\mathrm{T}}$.

The incompressibility constraint demands that 
\begin{equation}
\mathrm{det} \, \overline{\mathbb{F}} =
 [1
+\beta \omega   
\{ i  (\mathbf{a \cdot S})  e^{i \omega(\mathbf{S\cdot x}-t)}
+\mathrm{c.c.} \}]
( \lambda_{1}\lambda_{2}\lambda_{3})= 1,
\end{equation}
at all times. Therefore we must have
\begin{equation} \label{transverse}
\mathbf{a\cdot S}=0, \quad \lambda_{1}\lambda_{2}\lambda_{3}=1. 
\end{equation}

Also, $\check{J}=\mathrm{det} \, \check{\mathbb{F}}$ and 
$\overline{J}=\mathrm{det} \, \overline{\mathbb{F}}$ are given by 
\begin{equation} \label{Jbar} 
\check{J}=1, \quad \mathrm{and} \quad \overline{J}
=\check{J} \, J = J = \lambda_{1}\lambda_{2}\lambda_{3}=1.
\end{equation}

In the absence of body forces, the equations of motion,
written in the intermediate configuration, read
\begin{equation} \label{mtn2(B)} 
\mathrm{div}_{\mathbf{x}} \,
(\check{J}\, \overline{\mathbb{T}}
\check{\mathbb{F}}^{-1^{\mathrm{T}}}) =
\overline{\rho} \check{J} \,
\frac{\partial ^2 \overline{\mathbf{x}}}{\partial t^2}, \quad
\frac{\partial (\check{J} \, \overline{\mathbb{T}}
\check{\mathbb{F}}^{-1^{\mathrm{T}}})_{ij}} {\partial x_{j}} =
\overline{\rho}
\check{J} \, 
\frac{\partial ^2 \overline{\mathbf{x}}_i}{\partial t^2},
\end{equation} 
where $ \overline{\mathbb{T}}$ is the Cauchy stress tensor 
associated with  motion \eqref{xBar(B)}.
For a Mooney--Rivlin material, 
$\overline{\mathbb{T}}$ is related to the deformation gradient
through
\begin{equation} \label{Tbar(B)}
\overline{\mathbb{T}}= - \overline{p} \textbf{1}
+ C \overline{\mathbb{B}} - D \overline{\mathbb{B}}^{-1},
\end{equation}
where $ \overline{p}$ is the pressure.

 Upon using \eqref{Tbar(B)}, we have
\begin{equation} 
\check{J}\, \overline{\mathbb{T}}
\check{\mathbb{F}}^{-1^{\mathrm{T}}} =
 - \overline{p} \check{J}\, \check{\mathbb{F}}^{-1^{\mathrm{T}}}
+C \check{J}\, \check{\mathbb{F}}\mathbb{B}
-D  \check{J}\, \overline{\mathbb{B}}^{-1}
\check{\mathbb{F}}^{-1^{\mathrm{T}}}.
\end{equation} 
Hence, with equation \eqref{Jbar} and 
$ \partial(\check{J} \,  
\check{\mathbb{F}}^{-1^{\mathrm{T}}}_{ij})
/ \partial x_{j}=0$, 
the Euler--Jacobi--Piola identity, 
the equations of motion \eqref{mtn2(B)} reduce to
\begin{equation} \label{mnt3}
\check{\mathbb{F}}^{-1^{\mathrm{T}}}
\mathrm{grad}_{\textbf{x}}\overline{p}=
C  \mathrm{div}_{\textbf{x}}(\check{\mathbb{F}}\mathbb{B})
-D  \mathrm{div}_{\textbf{x}}( \overline{\mathbb{B}}^{-1}
\check{\mathbb{F}}^{-1^{\mathrm{T}}})
-\rho (\partial ^2 \overline{\mathbf{x}} /\partial t^2).
\end{equation}

Now we compute in turn the three terms of the right--hand side
of this equation.
For the first term, we have
\begin{align} \label{1stterm}
C  \mathrm{div}_{\textbf{x}}(\check{\mathbb{F}}\mathbb{B})
 & = C \beta \omega
\mathrm{div}_{\textbf{x}} [\mathbf{a} \otimes  
\{ i  e^{i \omega(\mathbf{S\cdot x}-t)}\mathbb{B}\mathbf{S} 
+\mathrm{c.c.} \} ]\nonumber \\
 & = - C  \beta \omega^2
\{  (\mathbf{S}\cdot \mathbb{B}\mathbf{S}) 
e^{i \omega(\mathbf{S\cdot x}-t)} 
+\mathrm{c.c.} \} \mathbf{a}
\end{align}

Using \eqref{transverse}, the second term  is found as
\begin{align} \label{2ndterm}
D  \mathrm{div}_{\textbf{x}}( \overline{\mathbb{B}}^{-1}
\check{\mathbb{F}}^{-1^{\mathrm{T}}}) = &
D \beta \omega^2
\{  [  (\mathbf{S\cdot S})\mathbb{B}^{-1}\mathbf{a}+
( \mathbf{a}\cdot \mathbb{B}^{-1}\mathbf{S}) \mathbf{S}] 
e^{i\omega(\mathbf{S\cdot x}-t)}
 +\mathrm{c.c.} \}  \nonumber \\
&+D \beta^2 \omega^3
( \mathbf{a}\cdot \mathbb{B}^{-1}\mathbf{a}) 
\{ i   (\mathbf{S\cdot S}) 
e^{2i\omega(\mathbf{S\cdot x}-t)} \mathbf{S}
 +\mathrm{c.c.} \} \nonumber \\
&+D \beta^2 \omega^3 
( \mathbf{a}\cdot \mathbb{B}^{-1}\mathbf{a}) 
e^{i\omega(\mathbf{S}- \mathbf{\widetilde{S}})\cdot \mathbf{x}} 
\{ i [(\mathbf{S}-\mathbf{\widetilde{S}})\cdot  \mathbf{\widetilde{S}}]
 \mathbf{S}
 +\mathrm{c.c.} \}.
\end{align}

Finally, the third term is given by 
\begin{equation} \label{3rdterm}
-\rho (\partial ^2 \overline{\mathbf{x}} /\partial t^2)=
\rho  \beta \omega^2 \mathbf{a}
\{ \mathbf{S} e^{i\omega(\mathbf{S\cdot x}-t)} +\mathrm{c.c.} \}.
\end{equation}

In view of \eqref{1stterm}, \eqref{2ndterm}, and \eqref{3rdterm}, 
we must take $\overline{p}$ (to within a cons\-tant term) of the form
\begin{equation}
\overline{p}=
 \beta \omega 
\{ i  p_1 e^{i\omega(\mathbf{S\cdot x}-t)} +\mathrm{c.c.} \}
 +\beta^2 \omega^2 
\{  p_2 e^{2i\omega(\mathbf{S\cdot x}-t)} +\mathrm{c.c.} \}
 + \beta^2 \omega^2 
p_3 e^{i\omega(\mathbf{S}- \mathbf{\widetilde{S}})\cdot \mathbf{x}},
\end{equation}
where $p_1, p_2, p_3$ are scalars ($p_3$ is real).
With this decomposition, we compute the right--hand side
of \eqref{mnt3} as
\begin{align} \label{rhs}
- \check{\mathbb{F}}^{-1^{\mathrm{T}}}
\mathrm{grad}_{\textbf{x}}\overline{p}= &
 \beta \omega^2
\{   p_1 e^{i\omega(\mathbf{S\cdot x}-t)}\mathbf{S} +\mathrm{c.c.} \}
 \nonumber \\
& -\beta^2 \omega^3 
\{ i  p_2 e^{2i\omega(\mathbf{S\cdot x}-t)}\mathbf{S} +\mathrm{c.c.} \}
\nonumber \\
& -\beta^2 \omega^3 
p_3 i(\mathbf{S}- \mathbf{\widetilde{S}})
e^{i\omega(\mathbf{S}- \mathbf{\widetilde{S}})\cdot \mathbf{x}}.
\end{align}

Now, using \eqref{1stterm}, \eqref{2ndterm}, 
\eqref{3rdterm} and \eqref{rhs}, we write the equations of motion
 \eqref{mnt3}, separating the respective coefficients of
$e^{i\omega(\mathbf{S\cdot x}-t)}$, 
$e^{2i\omega(\mathbf{S\cdot x}-t)}$, and 
$e^{i\omega(\mathbf{S}- \mathbf{\widetilde{S}})\cdot \mathbf{x}}$, as
\begin{equation} \label{motion}
\begin{array}{l}
-p_1 \mathbf{S} 
+ C ( \mathbf{S}\cdot \mathbb{B}\mathbf{S}) \mathbf{a}
+D[(\mathbf{S\cdot S})\mathbb{B}^{-1}\mathbf{a}
+( \mathbf{a}\cdot \mathbb{B}^{-1}\mathbf{S})\mathbf{S}]
=\rho \mathbf{a}, \\
-p_2 \mathbf{S} 
-D (\mathbf{a}\cdot \mathbb{B}^{-1}\mathbf{a})(\mathbf{S\cdot S})
\mathbf{S}=0,  \\
-p_3 (\mathbf{S} - \mathbf{\widetilde{S}})
+D (\mathbf{a}\cdot \mathbb{B}^{-1}\mathbf{a})
\{
[(\mathbf{S} - \mathbf{\widetilde{S}}) 
\cdot \mathbf{\widetilde{S}})]\mathbf{S}
+[(\mathbf{S} - \mathbf{\widetilde{S}}) \cdot \mathbf{S})]
\mathbf{\widetilde{S}}
\} 
=0.
\end{array}
\end{equation}

Writing \textbf{S} as $\mathbf{S}=\mathbf{S}^+ +i\mathbf{S}^-$,
 equation \eqref{motion}$_3$ is equivalent to 
\begin{equation}
[p_3 -2D (\mathbf{a}\cdot \mathbb{B}^{-1}\mathbf{a})
(\mathbf{S^-\cdot S^-})]\mathbf{S^-}=
[2D (\mathbf{a}\cdot \mathbb{B}^{-1}\mathbf{a})
(\mathbf{S^-\cdot S^+})]\mathbf{S^+}.
\end{equation}

Now, because the motion is inhomogeneous, $\mathbf{S}^-$
is not parallel to $\mathbf{S}^+$ and therefore, we must have
$p_3 -2D (\mathbf{a}\cdot \mathbb{B}^{-1}\mathbf{a})
(\mathbf{S^-\cdot S^-})=0$ and also
\begin{equation} \label{S+S-=0}
\mathbf{S^-\cdot S^+}=0,
\end{equation} 
which means that the planes of constant phase are 
orthogonal to the planes of constant amplitude.

%********************************************************************* 
\end{document}